\documentclass[prd, twocolumn, showpacs, amsmath, nofootinbib]{revtex4}

\begin{document}

\title{The Lanczos potential as a spin-2 field}
\author{Daniel Cartin}
\email{cartin@naps.edu}
\affiliation{Naval Academy Preparatory School\\
197 Elliot Street, Newport, RI 02841-1519}

\date{November 20, 2003}

\begin{abstract}

The Lanczos potential $L_{abc}$ acts as a tensor potential for the spin-2 field strength $W_{abcd}$ in an role similar to that of the vector potential $A_a$ for the Maxwell tensor $F_{ab}$. After some general considerations inspired by the example of electromagnetism, we consider the linear spin-2 theory and a Born-Infeld type action in terms of $L_{abc}$.

\end{abstract}


\maketitle

\section{Introduction}

The study of gravity, in the form of general relativity has been based traditionally on the metric tensor $g_{ab}$, and with good reason -- one can intuitively see how distances are altered by, for example, the passage of a gravitational wave or the proximity of a massive object. There are several other approaches that have been considered, such as twistor theory and the null surface formalism~\cite{null}, that are based on other mathematical objects. Although these are not widely used, they have provided insights into the broader theory, and can make explicit certain relations or results that are not as clear in the metric formulation.

With this in mind, we undertake the study of spin-2 theories based on the Lanczos potential $L_{abc}$~\cite{lan62}. This tensor acts as potential for the spin-2 field $W_{abcd}$, exactly as the 1-form $A_a$ does for the spin-1 field strength $F_{ab}$. In fact, we will see that there are similarities between the two that will allow us to take results from electromagnetism, and transfer them directly over to a corresponding spin-2 result. It is this analogy that we intend to study further, and develop the connections between electromagnetism, both the simplest and oldest of the gauge theories, and general relativity, well-known for its high degree of non-linearity and the difficulty in quantizing it.

Because the Lanczos potential for a Weyl candidate tensor $W_{abcd}$ (any tensor which possesses the symmetries of the Weyl tensor) is not widely known, we first give a brief explanation of its properties in Section \ref{desc}, and relate it to the linearized metric perturbation in Section \ref{compare}. This decomposition of $W_{abcd}$ in terms of $L_{abc}$ can only be done in four space-time dimensions (with arbitrary signature); as far as is known, there is no comparable result in higher dimensions~\cite{edg-hog02}. However, in four dimensions, we can write down the linearized spin-2 field equations in terms of $L_{abc}$, and obtain a correspondence between these solutions and those in terms of a metric perturbation $h_{ab}$.

Our eventual goal will be to have a Lanczos potential formulation that is equivalent to general relativity; as a first step, this paper looks at fields acting on flat space-time. In particular, we examine the Hamiltonian formulations of two actions -- the linear action in Section \ref{linear-sec}, the other based on the Born-Infeld action in Section \ref{BI-sec}. We also examine various gauge choices for the various fields obtained in the 3+1 decomposition of the tensor $L_{abc}$.

\subsection{The Lanczos potential}
\label{desc}

Based on work done by Lanczos~\cite{lan62}. and later updated by others (e.g. \cite{bam-cav83, lan-rev}), one can show that it is possible to express a Weyl candidate tensor $W_{abcd}$ in terms of a rank-3 tensor $L_{abc}$, now called the Lanczos potential. Starting with a generic $W_{abcd}$, we can write it in the form
\begin{equation}
\label{weyl-lanc}W_{abcd} = L_{ab[c;d]} +L_{cd[a;b]} - \!^* L^* _{ab[c;d]} - \!^* L^* _{cd[a;b]},
\end{equation}
where $L_{abc}$ is a potential chosen to have the symmetries
\begin{equation}
\label{L-sym}L_{abc} = - L_{bac} \qquad L_{[abc]} = 0,
\end{equation}
along with
\begin{subequations}
\label{gauge}
\begin{align}
\label{alg-gauge}     L_{ab}^{\ \ \ b} &= 0,
\\
\label{diff-gauge}
L_{abc} ^{\ \ \ \ ;c}&= 0.
\end{align}	
\end{subequations}
Working in terms of spinors, Illge~\cite{ill88} showed that any Weyl candidate spinor $W_{ABCD}$ = $W_{(ABCD)}$ can be written in terms of a potential $L_{ABCC'}$, such that\begin{equation}
\label{spin-Weyl-Lan}    W_{ABCD} = 2 \nabla^{X'} _{\ \ (A} L_{BCD)X'}.\end{equation}
This is the spinor form of the decomposition of the Weyl-like tensor $(\ref{weyl-lanc})$.The proof by Illge uses the spinor form of the Lanczos potential, and thus holds only for a four-dimensional Lorentzian spacetime; however, Bampi and Caviglia~\cite{bam-cav83} showed that this relation holds regardless of the signature of the metric (although still in four dimensions).

Condition $(\ref{alg-gauge})$ is known in the literature as the {\it Lanczos algebraic gauge}, while the condition $(\ref{diff-gauge})$ is called the {\it Lanczos differential gauge}. The term 'gauge' comes from the fact, proven by Bampi and Caviglia~\cite{bam-cav83}, that requiring these latter two conditions on $L_{abc}$ was not strictly necessary, and thus the trace and divergence of $L_{abc}$ could be fixed to be arbitrary tensors\footnote{There is also geometric reason to refer to these as gauge choices; see Hammon and Norris~\cite{ham-nor}.}. In this paper, we will assume that the algebraic gauge holds $(\ref{alg-gauge})$; since the Lanczos decomposition ($\ref{weyl-lanc}$) is invariant under transformations
\begin{equation}
\label{V-trans}
L_{abc} \to L_{abc} - \frac{2}{3} V_{[a} g_{b]c},
\end{equation}
there is no loss of generality in working with this gauge, by choosing $V_a = L_{ab} ^{\ \ \, b}$. In fact, if we suppose only the first three algebraic conditions $(\ref{L-sym})$ and $(\ref{alg-gauge})$ hold -- neglecting the differential relation $(\ref{diff-gauge})$ -- then the tensor $L_{abc}$ is equivalent to a spinor $L_{ABCC'} = L_{(ABC)C'}$\cite{zun75}. The differential gauge choice $(\ref{diff-gauge})$ is not the most natural to use, however, when working with the linear Hamiltonian; we explore this issue and look at other possibilities in Section \ref{gauge-con}.

As can be seen, the tensor $L_{abc}$ acts like an abelian connection for $W_{abcd}$; all of the usual non-linearity of the Levi-Civita connection is swept under the rug, so that there is no simple relationship between $L_{abc}$ and $\Gamma^a _{\  bc}$. One way to see how to quantify this relation is the following~\cite{lanc-kerr}, where we will use spinor notation. Suppose we consider a torsion-free metric connection $\nabla_{AA'}$; we can define a new connection ${\hat \nabla}_{AA'}$ by introducing a spinor $\Gamma_{ABCC'} = \Gamma_{(AB)CC'}$ such that
\begin{equation}
{\hat \nabla}_{AA'} \xi_B = \nabla_{AA'} \xi_B - 2 \Gamma^X _{\ BAA'} \xi_X.
\end{equation}
The new connection ${\hat \nabla}_{AA'}$ is still metric, but no longer torsion-free; the Weyl tensor ${\hat \Psi}_{ABCD}$ of this connection will be takes the form
\begin{equation*}
{\hat \Psi}_{ABCD} = \Psi_{ABCD} - 2 \nabla_{(A} ^{\ \ X'} \Gamma_{BCD)X'} - 4 \Gamma_{X(AB} ^{\ \ \ \ X'} \Gamma^X _{\ CD)X'}.
\end{equation*}
If we set ${\hat \Psi}_{ABCD} = 0$, then, using the spinor decomposition $(\ref{spin-Weyl-Lan})$ of $\Psi_{ABCD}$ in terms of the Lanczos spinor $L_{ABCC'}$, we have
\begin{equation}
\nabla^{X'} _{\ \ (A} L_{BCD)X'} = \nabla^{X'} _{\ \ (A} \Gamma_{BCD)X'} + 2 \Gamma_{X(AB} ^{\ \ \ \ X'} \Gamma^X _{\ CD)X'}.
\end{equation}
We have a non-linear relation between the Lanczos potential and the torsion necessary to insure a space-time has a vanishing Weyl tensor. Obviously, it is much easier if we are given the spinor $\Gamma_{ABCC'}$ to solve for the matching spinor $L_{ABCC'}$. However, it can be shown~\cite{lanc-kerr} that a particular choice of $L_{ABCC'} = \Gamma_{(ABC)C'}$ will work for Kerr-Schild space-times.
\subsection{Comparing linear spin-2 fields}
\label{compare}

Since we will be working with the Lanczos potential $L_{abc}$ as a potential for a spin-2 field, we next consider it relative to the more usual form of the linearized equations of motion based on a symmetric tensor $h_{ab}$. For our comparison, we start with the action as a function of $L_{abc}$,
\begin{equation}
\label{weak}{\cal L}_{lin} = \frac{1}{16} W^{abcd} W_{abcd},
\end{equation}
on a flat background, i.e. using the form $(\ref{weyl-lanc})$ of $W_{abcd}$ in terms of the tensor $L_{abc}$, but using partial derivatives instead of covariant derivatives, and we raise and lower indices with the flat metric $\eta_{ab}$. Varying with respect to the potential gives\begin{equation}\label{lin-H}C_{abcd} ^{\ \ \ \ ,d}= 0 \quad \Rightarrow \quad L_{abc,d} ^{\ \ \ \ \ ,d} = 0,\end{equation}where we have used the differential gauge $(\ref{diff-gauge})$ to get the second equation\footnote{On generic four-dimensional manifolds of any signature, the equation $\nabla^a C_{abcd} = 0$ give rise to the equation $\nabla^d \nabla_d L_{abc} = 0$ by the use of an algebraic identity between $L_{abc}$ and $C_{abcd}$~\cite{lan-rev}.}.

On the other hand, we can consider the spacetime metric to be of the
form $g_{ab} = \eta_{ab} + h_{ab}$ and take the linear order terms of the Einstein equations $R_{ab} = 0$. If we do so, then the spin-2 field equations are given by
\begin{equation}\label{lin-h}
h_{ab,c} ^{\ \ \ \ ,c} = 0,
\end{equation}
using the gauge condition $h_{ab} ^{\ \ ,b} = \frac{1}{2} h_{,a}$ where $h = \eta^{ab} h_{ab}$. The question now is how to relate these two equations of motion $(\ref{lin-H})$ and $(\ref{lin-h})$ for $L_{abc}$ and $h_{ab}$, respectively.

In fact, we can do this by writing $L_{abc}$ in terms of a rank-2 tensor $K_{ab}$, much as we did originally by expressing $W_{abcd}$ as a sum of derivatives of $L_{abc}$. The relevant result in this case is the following~\cite{and-edg}. Suppose we have a tensor with the symmetries of the
Lanczos potential; then, on a Ricci-flat spacetime with Lorentz signature 
metric $g_{ab}$, we can locally find a traceless, symmetric tensor $K_{ab}$,
such that\footnote{We assume here that we can choose a real, instead of complex, tensor $K_{ab}$. This can be done for several spacetimes of interest that are not necessarily Ricci-flat; for discussion and references, see Andersson and Edgar~\cite{and-edg}.}
\begin{equation}
\label{decomp1}
L_{abc} = - \nabla_{[a} K_{b]c} + \frac{1}{3} g_{c[a} \nabla^d K_{b]d}.
\end{equation}
In the specific case we are considering on a flat spacetime, by writing
\begin{equation*}
K_{ab} = \frac{1}{2} h_{ab} - \frac{1}{8} \eta_{ab} h,
\end{equation*}
then our general form $(\ref{decomp1})$ for $L_{abc}$ becomes\footnote{As an aside, we note that this relation illuminates the discussion in Section 5.7 of Penrose and Rindler~\cite{pen-rin}, and especially the origin of equations such as (5.7.12) in that work.}
\begin{equation}
\label{decomp2}
L_{abc} = \frac{1}{2} \biggl( h_{c[a,b]} - \frac{1}{6} \eta_{c[a} h_{,b]} \biggr).
\end{equation}
Using this last relation, we can now see the correspondence between solutions $L_{abc}$ of the equations of motion $(\ref{lin-H})$, and solutions $h_{ab}$ of the usual spin-2 equations $(\ref{lin-h})$, up to gauge transformations of $h_{ab}$. To show that $h_{ab}$ gives rise to a solution $L_{abc}$, we can  simply use the relation $(\ref{decomp2})$ to form the linearized Lanczos potential.  To go in the opposite direction, suppose that $h_{ab}$ and $h'_{ab}$ are two metric perturbations that give the same Lanczos potential $L_{abc}$.  Then, $\Delta h_{ab} \equiv h'_{ab} - h_{ab}$ gives a Lanczos potential of zero, and satisfies the linearized Einstein equations $R_{ab}(\Delta h_{cd}) = 0$.  Because of this, the Riemann tensor constructed from $\Delta h_{ab}$ is identically zero; hence, $\Delta h_{ab}$ corresponds to a constant multiple of a flat metric to linear order, which can be absorbed by a gauge transformation.

\section{General considerations}

We have seen that (at least at the level of linearized equations) we can relate the metric and Lanczos potential formalisms, mapping solutions of one type over to the other. The question is how to go beyond this to higher orders. One method would be to continue on a perturbative path, adding a second order term to the relation $(\ref{decomp1})$ given previously for $K_{ab}$, i.e.
\begin{equation}
K_{ab} ^{(2)} = \alpha_1 h_{ac} h^c _{\ b} + \alpha_2 h h_{ab} + O(h^3).
\end{equation}
Then we could continue along using a bootstrap method, order by order, with the action (\ref{weak}) and try to match the Einstein form of general relativity. There has also been other methods for using $L_{abc}$ as a spin-2 field, using a variety of action principles to obtain field equations (see~\cite{nov-net92, nov-nev} for some examples and further references).

However, in this paper, we will take a different approach by emphasizing the analogy to electromagnetism. Since the field strength $W_{abcd}$ of the Lanczos potential has the same symmetries as the Weyl tensor, it can described in terms of their electric and magnetic components $E_{ab}$ and $B_{ab}$, respectively, given as
\begin{equation}
E_{ac} = W_{abcd} t^b t^d \qquad B_{ac} = \,^* W_{abcd} t^b t^d,
\end{equation}
where $t^a$ is a unit time-like vector. So we can transfer over theoretical frameworks from electromagnetism into the study of spin-2 fields, when written in terms of $L_{abc}$. In the following, we will use this intuition to obtain several results about Lanczos potential actions.

\subsection{Equations of motion}

We begin by looking at the equations of motion for a general Lanczos potential action; with a Lagrangian ${\cal L} = {\cal L} (W_{abcd})$, we define
\begin{equation}
\label{G-def}
G^{abcd} = - 8 \frac{\delta {\cal L}}{\delta W_{abcd}}.
\end{equation}
Using the decomposition $(\ref{weyl-lanc})$ of $W_{abcd}$ in terms of $L_{abc}$, we can show that the equations of motion in this case are
\begin{equation}
\label{EOM}
\nabla^a G_{abcd} = 0.
\end{equation}
Note that, unlike the case of electromagnetism, there is no separate Bianchi identity; the divergence of the Weyl tensor comes directly from the Bianchi identity $\nabla_{[a} R_{bc]de} = 0$, and is given by
\begin{equation}
\label{Weyl-div}
\nabla^a C_{abcd} = \nabla_{[c} R_{d]b} - \frac{1}{6} g_{b[c} \nabla_{d]} R.
\end{equation}
Next, we write down the 3+1 decomposition of the equations of motion on a flat background, splitting four-dimensional spacetime into its space and time portions and replacing covariant derivatives $\nabla_a$ with partial derivatives. We also split the four-dimensional metric into $g_{ab} = q_{ab} - t_a t_b$, with $t^a$ a constant timelike vector ($t^a t_a = -1$) and $q^{ab}$ the spatial metric. Since the equations feature $W_{abcd}$ only as it appears in $G_{abcd}$, we use the fields
\begin{equation}
\label{D-H}
D_{ac} = G_{abcd} t^b t^d \qquad H_{ac} = \,^* G_{abcd} t^b t^d,
\end{equation}
Then, the field equations (\ref{EOM}) become
\begin{equation}
\label{max-1}
\frac{\partial H^{ab}}{\partial t} + \epsilon^{ij(a} \partial_i D_j ^{\ b)} = 0  \qquad
\nabla^a H_{ab} = 0,
\end{equation}
and
\begin{equation}
\label{max-2}
\frac{\partial D^{ab}}{\partial t}  - \epsilon^{ij(a} \partial_i H_j ^{\ b)} = 0.
\qquad \nabla^a D_{ab} = 0
\end{equation}
For electromagnetism, the equations $(\ref{max-1})$ are a geometric restriction on the fields from the Bianchi identity for $F_{ab}$; here, they are field equations, and thus are satisfied {\it only} on-shell.

One way to obtain solutions for the fields $D_{ab}$ and $H_{ab}$ is by linearizing the Weyl tensor $C_{abcd}$. Suppose we start with a metric $g_{ab}$ whose Weyl tensor has zero divergence (i.e. a C-space), its Levi-Civita connection $\nabla_a$ and its Weyl tensor $C_{abcd}$, and linearize the metric around the flat metric $\eta_{ab}$; expanding in terms of a parameter $\epsilon$, we have
\begin{align}
\begin{split}
\nabla_a &= \partial_a + \epsilon \nabla_a ^{(1)} + O(\epsilon^2),
\\
C_{abcd} &= \epsilon \, C^{(1)} _{abcd} + O(\epsilon^2).
\end{split}
\end{align}
There is no Weyl tensor for the flat metric $\eta_{ab}$; the potential $L^{(0)} _{abc}$ is pure gauge. Therefore, to order $\epsilon$, the Bianchi identity $(\ref{Weyl-div})$ for the linearized Weyl tensor is exactly the same as the equations of motion for the Lanczos potential $(\ref{EOM})$. Specifically,
\begin{equation}
0 = \nabla^a C_{abcd} = \epsilon \partial^a C_{abcd} ^{(1)} + O(\epsilon^2).
\end{equation}
From this correspondence, we can find solutions to our equations of motion, based on linearization of the Weyl tensor.

Of course, once we have a solution for $D_{ab}$ and $H_{ab}$ (and hence $E_{ab}$ and $B_{ab}$, based on the particular action ${\cal L}$ we are using), we need to find how other matter is affected by this. Just as electromagnetism has its Lorentz force law, we conjecture that test particles would have trajectories given by a something similar to the geodesic deviation equation. We suppose there are test particles to be traveling along a family of trajectories, with each particle having a tangent vector $u^a$ to its path, and a vector $n^a$ which gives the separation between the particles. Then the interaction between the particles and the field strength $W_{abcd}$ is assumed to be
\begin{equation}
\frac{d^2 n^a}{d \sigma^2} + W ^a _{\ \ bcd} u^b n^c u^d = 0,
\end{equation}
where $\sigma$ is the parameter along the geodesics. However, this only seems applicable to point particles, where we can think in terms of geodesics.

For a more general relation between $L_{abc}$ and generic matter fields, we can introduce a coupling to a matter current of the form
\begin{equation}
\label{matter-action}
{\cal L}_{int} = L_{abc} \, J^{abc},
\end{equation}
where $J^{abc}$ is the matter current. We can relate this to a symmetric tensor $T^{ab}$, just as we could write the potential $L_{abc}$ in terms of a tensor $K_{ab}$, by again using (\ref{decomp1}).
Choosing
\[
K^{ab} = T^{ab} - \frac{1}{4} \eta^{ab} T,
\]
with $T$ the trace of $T^{ab}$, we have that
\begin{equation}
J^{abc} = T^{c[a,b]} - \frac{1}{3} \eta^{c[a} T^{,b]},
\end{equation}
and the equations of motion become
\begin{equation}\label{current}
C^{abcd}_{\ \ \ \ ,d} = J^{abc} = T^{c[a,b]} - \frac{1}{3} \eta^{c[a} T^{,b]}.
\end{equation}
We compare this to the divergence of the Weyl tensor $(\ref{Weyl-div})$; using the Einstein equations $R_{ab} - \frac{1}{2} g_{ab} R = T_{ab}$, where $T_{ab}$ is the usual energy-momentum tensor for matter, leads to the identification of the tensor $T^{ab}$ as the energy-momentum tensor. The fact that the divergence of $T^{ab}$ is zero comes from the trace-free nature of the left-hand side of $(\ref{current})$.

\subsection{Duality invariance}

In electromagnetism, only the linear Maxwell equations are invariant under the duality transformation $E_a \to B_a, B_a \to - E_a$. For more general actions, other types of duality have to be considered. One possibility is to have rotations between $E_a, B_a$ and $D_a, H_a$, namely,
\begin{align}
\begin{split}
\label{Max-dual}
E_a + i H_a &\to e^{i \theta} (E_a + i H_a),
\\
D_a + i B_a &\to e^{i \theta} (D_a + i B_a),
\end{split}
\end{align}
rotating through a given angle $\theta$. We will do something comparable, following the procedure used by Gibbons and Rasheed~\cite{gib-ras95} for electromagnetism, and consider infinitesimal transformations of the form
\begin{align}
\label{xform}
\begin{split}
\delta F_{abcd} = \,^* G_{abcd},
\\
\delta G_{abcd} = \:^*F_{abcd},
\end{split}
\end{align}
Substituting in our defining relation $(\ref{G-def})$ for $G_{abcd}$, these infinitesimal transformations $(\ref{xform})$ give
\begin{equation*}
\frac{1}{2} \epsilon^{ab} _{\ \ ef} W^{efcd} = \frac{1}{2} \epsilon_{ijrs} G^{rs} _{\ \ kl} \frac{\partial}{\partial W_{ijkl}} \biggl( -8 \frac{\partial {\cal L}}{\partial W_{abcd}} \biggr).
\end{equation*}
Using the commutativity of the derivatives, and substituting the definition of $G_{abcd}$ into this equation, we get
\begin{equation*}
\frac{1}{2} \epsilon^{ab} _{\ \ ef} W^{efcd} = 32 \epsilon_{ijrs} g_{km} g_{ln} \frac{\partial {\cal L}}{\partial W_{rsmn}} \frac{\partial}{\partial W_{abcd}} \biggl( \frac{\partial {\cal L}}{\partial W_{ijkl}} \biggr).
\end{equation*}
Because of the pair symmetry $\epsilon_{ijkl} = \epsilon_{klij}$, we can re-write this as
\begin{equation*}
\frac{1}{2} \epsilon^{ab} _{\ \ ef} W^{efcd} =  16 \epsilon_{ijrs} g_{km} g_{ln} \frac{\partial}{\partial W_{abcd}} \biggl( \frac{\partial {\cal L}}{\partial W_{rsmn}} \frac{\partial {\cal L}}{\partial W_{ijkl}} .\biggr)\end{equation*}
If we integrate this equation with respect to $W_{abcd}$, we get the following relation
\begin{equation*}
\frac{1}{2} \epsilon^{ab} _{\ \ ef} W^{efcd} W_{abcd} = 32 \epsilon_{ijrs} g_{km} g_{ln} \biggl( \frac{\partial {\cal L}}{\partial W_{rsmn}} \frac{\partial {\cal L}}{\partial W_{ijkl}} \biggr).
\end{equation*}
Notice that we have set the integration constant equal to zero, to ensure that our weak-field Lagrangian $(\ref{weak})$ satisfies the equation. Finally, substituting into this equation, using the definition of $G^{abcd}$, we find that
\begin{equation}
\label{dual-rel}
W^{abcd} W^*_{abcd} = G^{abcd} G^* _{abcd},
\end{equation}
i.e. $E^{ab} B_{ab} = D^{ab} H_{ab}$. We can see that this relation is obeyed by our Maxwell-type action for $W_{abcd}$ $(\ref{weak})$, since $D_{ab} = E_{ab}$ and $H_{ab} = B_{ab}$.

Not only is there the possibility of duality between the electric and magnetic fields, but also a Hodge duality involving the potential $L_{abc}$ itself. Suppose we define a tensor $G_{abc}$ such that $M_{abc} = \!^* L_{abc}$. Notice that $M_{abc}$ has exactly the same symmetries as $H_{abc}$, since
\begin{equation}
L_{[abc]} = 0 \Leftrightarrow M_{ab} ^{\ \ b} = 0
\end{equation}
and vice versa. This establishes a duality between the electric and magnetic fields of the two potentials $L_{abc}$ and $M_{abc}$, since
\begin{equation}
E^M _{ab} = B^L _{ab} \qquad B^M _{ab} = - E^L _{ab},
\end{equation}
where the superscript refers to the potential used in the definition of each field. If we let
\begin{equation}
\label{linear-duality}
(L_{abc} + iM_{abc}) \to e^{i \alpha} (L_{abc} + iM_{abc}),
\end{equation}
this introduces a corresponding transformation of the electric and magnetic fields $(E_{ab} + iB_{ab}) \to e^{i \alpha} (E_{ab} + iB_{ab})$. However, only the linear spin-2 equations are invariant under this transformation; for the more general case, we would want to consider transformations similar to the duality of the electromagnetic fields $(\ref{Max-dual})$, except in terms of the potentials. Since $G_{abcd}$ has the same symmetries as the Weyl tensor, it has its own Lanczos-type potential, which we will call $P_{abc}$\footnote{In other words, $E^P _{ab} = D_{ab}$ and $B^P _{ab} = H_{ab}$.}. Then transformations of the form
\begin{align}
\begin{split}
(L_{abc} + i^* \! P_{abc}) & \to e^{i \alpha} (L_{abc} + i^* \! P_{abc}),
\\
(M_{abc} + i P_{abc}) & \to e^{i \alpha} (M_{abc} + i P_{abc}),
\end{split}
\end{align}
will lead to duality relations, analogous to those in $(\ref{Max-dual})$, between the fields $E_{ab}$ and $H_{ab}$, and between $B_{ab}$ and $D_{ab}$.

\section{Linear spin-2 action}
\label{linear-sec}

As we saw in Section \ref{compare}, we can relate the equations of motion for linear perturbations $h_{ab}$ and $L_{abc}$; in this section, we look at the Hamiltonian formulation of the same action.
Again we take the case of a flat space-time with a constant time vector $t^a$; we save for a future paper the case of more general background space-times, and the question of what constraints may exist for these backgrounds to ensure consistency.

\subsection{Hamiltonian formalism}

Here again, we split the metric $g_{ab}$ into its spatial part $q_{ab}$ and time-like vector $t^a$; after this is done, we define the potentials
\begin{subequations}
\label{pot}
\begin{align}
\label{pot-phi}
\phi_{ab} &= L^{c(ij)} t_c q_{ai} q_{bj},
\\
\label{pot-A}
A_{abc} &= L^{ijk} (q_{ai} q_{bj} q_{ck} - q_{i[a} q_{b]c} q_{jk}),
\end{align}
\end{subequations}
and
\begin{equation}
\label{pot-gauge}
\chi_{ab} = L^{c[ij]} t_c q_{ai} q_{bj} \qquad V_a = L^{ijk} t_i q_{aj} t_k.
\end{equation}
We can show that $A_{abc}$ retains the symmetries of $H_{abc}$, namely
\begin{equation}
A_{abc} = A_{[ab]c} \qquad A_{[abc]} = 0
\end{equation}
In addition, the algebraic gauge condition $(\ref{alg-gauge})$ for the tensor $L_{abc}$ makes $\phi_{ab}$ and $A_{abc}$ trace-free:
\begin{equation}
\phi_{ab} q^{ab} = A_{abc} q^{bc} = 0.
\end{equation}
As will be seen, the Lanczos potential is made up of two spin-2 fields, $\phi_{ab}$ and $A_{abc}$, and two gauge fields $V_a$ and $\chi_{ab}$ with three degrees of freedom each. After making the appropriate gauge choice, only one of these will remain a dynamical field; we will return to this subject in Section \ref{gauge-con}.

The electric and magnetic parts of the field strength $W_{abcd}$ are given in terms of the potentials $(\ref{pot})$ and $(\ref{pot-gauge})$ by
\begin{subequations}
\label{EB}
\begin{align}
E_{ab} & = - {\dot \phi_{ab}} + \frac{3}{2} \partial_{(a} V_{b)} - \frac{1}{2} q_{ab} \partial^c V_c - \partial^c A_{c(ab)},
\\
\label{B-to-S}
B_{ab} &= \epsilon_{ij(a} S^{ij} _{\ \ b)},
\end{align}
\end{subequations}
where we have defined $\epsilon_{abc} = \epsilon_{abcd} t^d$ and
\begin{align}
\label{S-def}
\begin{split}
S^{ij} _{\ \ b} &= \frac{1}{2} {\dot A^{ij} _{\ \ b}} + \partial_b \chi^{ij} + \partial^{[i} \phi_b ^{\ j]} + \partial^{[i} \chi_b ^{\ j]} 
\\
&\ + \frac{1}{2} \partial^a \phi_a ^{\ [i} q^{j]} _{\ \ b} + \frac{3}{2} \partial^a \chi_a ^{\ [i} q^{j]} _{\ \ b}.
\end{split}
\end{align}
Because $E_{ab}$ and $B_{ab}$ are contractions of $W_{abcd}$ with two time vectors $t^a$ (instead of the single vector used with electromagnetism), both $E_{ab}$ and $B_{ab}$ contain the time derivative of a potential.

We start with the linear spin-2 action ($\ref{weak}$), written in terms of the electric and magnetic fields as
\begin{equation}{\cal L}_{lin} =  \frac{1}{2} (E^{ab} E_{ab} - B^{ab} B_{ab}).
\end{equation}
If we use the more appropriate tensor $S_{abc}$, then from the definition of the magnetic field $B_{ab}$ in terms of $S_{abc}$, given in ($\ref{B-to-S}$),
\begin{equation}{\cal L}_{lin} = \frac{1}{2} E^{ab} E_{ab} - S^{abc} S_{abc}.
\end{equation}
Since
\begin{equation}
\frac{{\cal L}_{lin}}{\delta E_{ab}} = - \frac{{\cal L}_{lin}}{\delta {\dot \phi_{ab}}} \qquad \frac{{\cal L}_{lin}}{\delta S_{abc}} = \frac{1}{2} \frac{{\cal L}_{lin}}{\delta {\dot A_{abc}}},
\end{equation}
we have that the momenta for this action are
\begin{equation}
\pi^{ab} = - E^{ab} \qquad \pi^{abc} = - S^{abc}.
\end{equation}
From this, the linear spin-2 Hamiltonian is
\begin{align}
\label{lin-ham}
\begin{split}
{\cal H}_{lin} &= \frac{1}{2} \pi^{ab} \pi_{ab} - \pi^{abc} \pi_{abc} - \pi^{ab} \partial^k A_{kab}
\\
&\  - 2 \pi^{abc} \partial_a \phi_{bc} - 3 \pi^{abc} \partial_c \chi_{ab}  + \frac{3}{2} \pi^{ab} \partial_a V_b.
\end{split}
\end{align}
The time derivatives of both $V_a$ and $\chi_{ab}$ are absent from the Hamiltonian, so
we have that
\begin{subequations}
\begin{align}
\frac{\delta {\cal H}_{lin}}{\delta \chi_{ab}} &= 3 \partial_c \pi^{abc} = 0,
\\
\frac{\delta {\cal H}_{lin}}{\delta V_a} &= - \frac{3}{2} \partial_b \pi^{ab} = 0.
\end{align}
\end{subequations}
This gives use two constraints,
\begin{equation}
C^{ab} = \partial_c \pi^{abc} \qquad C^{a} = \partial_b \pi^{ab}.
\end{equation}
By using the form of the Hamiltonian ($\ref{lin-ham}$), we can show that these commute with ${\cal H}_{lin}$ and each other.

For the time derivatives of our fields $\phi_{ab}$ and $A_{abc}$, the equations of motion are
\begin{subequations}
\begin{align}
\label{pi-phi-def}
{\dot \phi^{ab}} &= \pi^{ab} + \frac{3}{2} \partial^{(a} V^{b)} - \frac{1}{2} q^{ab} \partial_c V^c - \partial_c A^{c(ab)},
\\
\label{pi-A-def}
\begin{split}
{\dot A^{abc}} &= - 2 \pi^{abc} - 2 \partial^c \chi^{ab} + 2 \partial^{[a} \chi^{b]c} - 2 \partial^{[a} \phi^{b]c}
\\
&\ - (\partial_d \phi^{d[a} ) q^{b]c} - 3 (\partial_d \chi^{d[a} ) q^{b]c}.
\end{split}
\end{align}
\end{subequations}
Note that, in writing these equations, we have added terms that preserve the symmetries of $\phi_{ab}$ and $A_{abc}$. Finally, we have that
\begin{subequations}
\begin{align}
\label{phi-dot}
{\dot \pi^{ab}} &= - 2 \partial_c \pi^{c(ab)},
\\
\label{A-dot}
{\dot \pi^{abc}} &= - \partial^{[a} \pi^{b]c}.
\end{align}
\end{subequations}
By combining these equations together, we expect to get a wave equation for the dynamical field; this requires choosing a proper set of gauge conditions, which we consider next.

\subsection{Gauge conditions}
\label{gauge-con}

When we plug in the expressions (\ref{pi-phi-def}) and (\ref{pi-A-def}) for $\pi^{ab}$ and $\pi^{abc}$, respectively, in terms of the various fields into the equations of motion (\ref{phi-dot}) and (\ref{A-dot}), we find that
\begin{subequations}
\begin{align}
0 &= ({\ddot  \phi}^{ab} - \partial_c \partial^c \phi^{ab}) - \frac{3}{2} \partial^{(a} T^{b)} + \frac{1}{2} q^{ab} \partial_c T^c,
\\
\begin{split}
0 &= ({\ddot  A}^{ab} - \partial_d \partial^d A^{abc}) + 2 \partial^c T^{ab}
\\
&\ - 2 \partial^{[a} T^{b]c} + 3(\partial_d T^{d[a}) q^{b]c},
\end{split}
\end{align}
\end{subequations}
where
\begin{subequations}
\begin{align}
T^a &= {\dot V}^a - \partial_b \chi^{ab} - \partial_b \phi^{ab},
\\
T^{ab} &= {\dot \chi}^{ab} + \frac{1}{2} \partial^{[a} V^{b]} + \frac{1}{2} \partial_c A^{abc}.
\end{align}
\end{subequations}
Doing the same for the constraint relations give conditions on the mixed derivatives $\partial_b {\dot \phi}^{ab}$ and $\partial_c {\dot A}^{abc}$, given by
\begin{subequations}
\begin{align}
\label{phi-div}
\partial_b {\dot \phi}^{ab} &= \frac{3}{4} \partial_c \partial^c V^a + \frac{1}{4} \partial^a \partial^c V_c + \frac{1}{2} \partial_i \partial_j A^{aij},
\\
\label{A-div-anti}
\partial_c {\dot A}^{abc} &= - 2 \partial_c \partial^c \chi^{ab} - \partial_c \partial^{[a} \chi^{b]c} - \partial_c \partial^{[a} \phi^{b]c}.
\end{align}
\end{subequations}
At this stage, we have the possibility of wave equations for both $\phi^{ab}$ and $A^{abc}$; since we want to limit this to only one dynamical spin-2 field, we must make the appropriate gauge choice. As mentioned in Section \ref{desc}, we are free to choose the Lanczos differential gauge (\ref{diff-gauge}); if we break it into its 3+1 components, we find that
\begin{subequations}
\begin{align}
t^a q^{bk} L_{abc} ^{\ \ \ \ ,c} = 0 &\Rightarrow {\dot V}^k = \partial_i (\phi^{ik} + \chi^{ik}),
\\
q^{aj} q^{bk} L_{abc} ^{\ \ \ \ ,c} = 0 &\Rightarrow {\dot \chi}^{jk} = - \frac{1}{2} \partial^{[j} V^{k]} - \frac{1}{2} \partial_i A^{jki}.
\end{align}
\end{subequations}
Thus, taking this differential gauge in its entirety will give $T^a = T^{ab} = 0$, but will leave us still with two dynamical fields.

We have written the decomposition of the Lanczos differential gauge above in a manner to suggest our eventual solution. If we take only half the gauge condition, then we get a wave equation for one of the fields. After this is done, we can choose an additional gauge condition to reduce the degrees of freedom to one massless spin-2 field. We have already mentioned that the field tensor $W_{abcd}$ is invariant under transformations by a vector $V_a$, given by $(\ref{V-trans})$. However, on any conformally flat space-time, $W_{abcd}$ is also invariant under the transformation~\cite{nov-net92}
\begin{equation}
L_{abc} \to L_{abc} + F_{ab;c} - F_{c[a;b]},
\end{equation}
where $F_{ab}$ is an anti-symmetric tensor. By using this additional gauge freedom, we can make an appropriate choice of gauge.

In addition, because we are dealing with the linear spin-2 field, we have the additional symmetry (\ref{linear-duality}) between the potential $L_{abc}$ and its Hodge dual $M_{abc} =\,^* L_{abc}$. Thus, for whatever gauge choice we make, when $\phi^L _{ab}$ is the only dynamical field for the potential $L_{abc}$, then $A^M _{abc}$ will be the same for the potential $M_{abc}$. This can already be seen in the Lanczos differential gauge, since
\[
q^{aj} q^{bk} L_{abc} ^{\ \ \ \ ,c} = 0\Leftrightarrow T^{ab} = 0,
\]
while
\[
q^{aj} q^{bk} M_{abc} ^{\ \ \ \ ,c} = q^{aj} q^{bk \ *} L_{abc} ^{\ \ \ \ ,c} = 0 \Leftrightarrow T^{a} = 0.
\]
We will see that the gauge condition\footnote{This condition is used by Novello and Neves~\cite{nov-nev} to isolate one of the two spin-2 fields as dynamical. They consider this as a necessary and sufficient condition which does not allow for the dual choice, so that what we call $\phi_{ab}$ is their only dynamical field.}
\begin{equation}
\,^* L^{a(bc)} _{\quad \ \ ,a} = 0
\end{equation} 
will give the Hamiltonian formulation in terms of the field $\phi_{ab}$; the dual condition
\begin{equation}
L^{a(bc)} _{\quad \ \ ,a} = 0
\end{equation} 
will do the same in terms of $A_{abc}$.

If we choose the gauge condition $\!^* L^{a(bc)} _{\quad \ \ ,a} = 0$, this is equivalent to the relations
\begin{subequations}
\begin{align}
\label{curl-chi}
\partial_{[a} \chi_{bc]} &= 0,
\\
\begin{split}
{\dot A_{abc}} &= 2 \partial_{[a} \phi_{b]c} + 2 \partial_{[a} \chi_{b]c}
\\
&\ + \partial_k \phi_{[a} ^{\ \ k} q_{b]c} + \partial_k \chi_{[a} ^{\ \ k} q_{b]c},
\end{split}
\\
\label{chi-dot}
{\dot \chi^{ab}} &= \frac{1}{2} \partial_c A^{abc} - \frac{3}{2} \partial^{[a} V^{b]},
\end{align}
\end{subequations}
In the last relation (\ref{chi-dot}), we have used the identity
\begin{equation}
\partial_{[a} A_{bc]d} = (\partial_k A_{[ab} ^{\ \ \ k}) q_{c]d}.
\end{equation}
We have also taken advantage of the relation $\partial_d \partial^{[a} A^{bc]d} = 0$, which results from the gauge equations (\ref{curl-chi}) and (\ref{chi-dot}).
When this gauge is chosen, the field equation (\ref{A-dot}) for ${\dot \pi}^{ab}$ becomes a constraint on the mixed derivative of $A^{c(ab)}$,
\begin{equation}
\partial_c {\dot A}^{c(ab)} = \frac{3}{2} \partial_c \partial^{(a} \chi^{b)c} + \partial_c \partial^c \phi^{ab} - \frac{5}{2} \partial_c \partial^{(a} \phi^{b)c}.
\end{equation}
This, along with the previous condition (\ref{A-div-anti}) on $\partial_c A^{abc} = -2 \partial_c A^{c[ab]}$ completely specifies the divergence of ${\dot A}_{abc}$.

On the other hand, choosing the condition $L^{a(bc)} _{\quad \ \ ,a} = 0$ gives the equations
\begin{subequations}
\begin{align}
\label{div-V}
\partial^a V_a &= 0,
\\
{\dot \phi}^{ab} &= \partial_c A^{c(ab)} + \frac{1}{2} \partial^{(a} V^{b)},
\\
\label{V-dot}
{\dot V}^a &= \partial_b (3 \chi^{ab} - \phi^{ab}).
\end{align}
\end{subequations}
This choice reduces the equation of motion (\ref{phi-dot}) for ${\dot \pi}^{abc}$ to a constraint on $\phi^{ab}$ given by
\begin{align}
\begin{split}
\partial^{[a} {\dot \phi}^{b]c} &= \frac{5}{2} \partial^{[a} \partial_d A^{b]cd} + \frac{3}{4} \partial^c \partial^{[a} V^{b]}
\\
&\ + (\partial^i \partial^j A^{[a} _{\ \ \ ij}) q^{b]c} - \frac{1}{2} (\partial^i \partial^j V^{[a}) q^{b]c},
\end{split}
\end{align}
where we have used the fact that the gauge equations (\ref{div-V}) and (\ref{V-dot}) imply that $\partial_c \partial_d \phi^{cd} = 0$. Together with the constraint (\ref{phi-div}), we have conditions on both the curl and the divergence of the time derivative ${\dot \phi}^{ab}$.

\section{Analogue of the Born-Infeld action}
\label{BI-sec}

Now that we have examined the linear spin-2 theory in terms of the Lanczos potential, the question is how to proceed to higher order. Since there are only two cubic invariants for Weyl-like tensors, one could study an action of the form
\begin{align*}
{\cal L} &= \frac{1}{16} W^{abcd} W_{abcd} + \alpha_1 W_{abcd} W^{abrs} W^{cd} _{\ \ \ rs}
\\
&\ + \alpha_2 ^{\ *} W_{abcd} W^{abrs} W^{cd} _{\ \ \ rs} + O(W^4),
\end{align*}
with two coefficients $\alpha_i$ to be determined. Then, one can look at the 3+1 decomposition, and see if there is a choice of coefficients that will give equations for $\phi^{ab}$ that match those of the Einstein equations to the same order. Yet this would require the same effort at every perturbation order, without a guiding principle to fix the coefficients.

To develop the non-linear theory in a more systematic way, we continue our analogy with spin-1 Maxwell theory, and its variations. One of the more well-known generalizations of the standard action for electromagnetism is that of Born and Infeld,
\begin{equation}
\label{BI}
S = \int \sqrt{g} \ \sqrt{1 + \frac{1}{2} F^{ab} F_{ab} - \frac{1}{16} \bigl(F^{ab} F^* _{ab} \bigr)^2},
\end{equation}
which can be written in a simpler form as
\begin{equation}
S = \int \sqrt{ \det (g_{ab} + F_{ab})}.
\end{equation}
This action has been re-examined a great deal recently, due to its connections with string theory (see, e.g.,~\cite{string-BI}). In addition, there has also been research which studies whether a gravitational version of Born-Infeld theory can be developed~\cite{grav-BI}. Thus, it is a natural step to look at the Lanczos potential version of the Born-Infeld action as a choice for a non-linear spin-2 theory.

Unlike the field strength $W_{abcd}$, both the metric $g_{ab}$ and the Maxwell tensor $F_{ab}$ can be taken as matrices, and we can find their determinant. However, $W_{abcd}$ is a bivector, and so we can treat it as a matrix of matrices. Then, if we trace over the "internal" matrices (the first pair of indices), and take the determinant of the second pair of indices, we can arrive at an expression similar to the Born-Infeld action $(\ref{BI})$. So, defining
\begin{equation}
\label{delta}
\Delta = - \frac{g}{4!} \epsilon^{abcd} \epsilon^{ijkl} M^p _{\ qai} M^q _{\ rbj} M^r _{\ sck} M^s _{\ pdl},
\end{equation}
where $g$ is the determinant of the metric $g_{ab}$, and
\begin{equation}
M^a _{\ bcd} = \delta^a _b g_{cd} + W^a _{\ bcd},
\end{equation}
we take as our spin-2 action ${\cal L}_{BI} = \frac{1}{2} \Delta^{1/2}$. Computing this in terms of the tensor $W^{abcd}$, we find that the action is of the form
\begin{equation*}
{\cal L}_{BI} = \sqrt{g} \sqrt{1 + \frac{1}{8} W^{abcd} W_{abcd} - \frac{1}{256} \bigl(W^{abcd} W^* _{abcd} \bigr)^2 }.
\end{equation*}
Written in terms of the fields $E_{ab}$ and $B_{ab}$, this is
\begin{equation}
{\cal L}_{BI} = \sqrt{g} \sqrt{1 + (E^{ab} E_{ab} - B^{ab} B_{ab}) - (E^{ab} B_{ab})^2}.
\end{equation}
Thus, our equations of motion are given by the Maxwell-type equations $(\ref{max-1}$) and ($\ref{max-2}$), where the fields $D_{ab}$ and $H_{ab}$ are are now
\begin{subequations}
\begin{align}
D_{ab} &= \frac{- E_{ab} + B_{ab} (E^{ab} B_{ab})}{\sqrt{1 + E^{ab} E_{ab} - B^{ab} B_{ab} - (E^{ab} B_{ab})^2}},
\\
H_{ab} &= \frac{- B_{ab} - E_{ab} (E^{ab} B_{ab})}{\sqrt{1 + E^{ab} E_{ab} - B^{ab} B_{ab} - (E^{ab} B_{ab})^2}}.
\end{align}
\end{subequations}
The difference in signs between those in Lagrangian ${\cal L}_{BI}$ come from the definition of $D_{ab}$ and $H_{ab}$ in terms of $G_{abcd}$, from (\ref{G-def}) and (\ref{D-H}). These fields are related to the momenta of $\phi^{ab}$ and $A^{abc}$ by
\begin{equation}
\label{BI-mom}
\pi^{ab} =  D^{ab} \qquad \pi^{abc} = \frac{1}{2} \epsilon^{ab} _{\ \ \ i} H^{ic}.
\end{equation}

If we write the duality relation (\ref{dual-rel}) in terms of the momenta, we find that
\begin{equation}
\epsilon_{ij} ^{\ \ \ a} \pi_{ab} \pi^{ijb} = E^{ab} B_{ab};
\end{equation}
from this, the Hamiltonian can be written as
\begin{align}
\begin{split}
{\cal H}_{BI} &= - \sqrt{1 + 2 \pi^{abc} \pi_{abc} - \pi^{ab} \pi_{ab} - (\epsilon_{ij} ^{\ \ \ a} \pi_{ab} \pi^{ijb})^2} 
\\
&\  - \pi^{ab} \partial^k A_{kab} - 2 \pi^{abc} \partial_a \phi_{bc} - 3 \pi^{abc} \partial_c \chi_{ab} 
\\
&\  + \frac{3}{2} \pi^{ab} \partial_a V_b.
\end{split}
\end{align}
This Hamiltonian has two differences with the one seen in non-linear electromagnetism. First, the expression under the square root does not appear to be positive definite, which may limit its utility. Requiring that the expression under the square root is non-negative may restrict the solutions of the field equations. Second, because the fields $\phi^{ab}$ and $A^{abc}$ appear exactly as they did in the linear action, the equations of motion and the constraints are the same as before. The only difference is that the momenta $\pi^{ab}$ and $\pi^{abc}$ are defined in (\ref{BI-mom}); obviously these reduce to those of the linear case when considering weak fields.

As a final note, we comment on the ease of adding in an electromagnetic field into our non-linear action; by defining
\begin{equation}
{\tilde M}^a _{\ bcd} = \delta^a _b (g_{cd} + F_{cd}) + W^a _{\ bcd},
\end{equation}
and substituting this into our relation $(\ref{delta})$ for $\Delta$, we obtain an action which includes interaction terms between the electromagnetic and Lanczos field strengths. For other types of matter, one would have to add in an interaction term as discussed previously.

\end{document}